# Invariants and conservation laws of physical quantities in Minkowski space

Yuriy A. Spirichev

The State Atomic Energy Corporation ROSATOM, "Research and Design Institute of Radio-Electronic Engineering" - branch of joint-stock company "Federal Scientific-Production Center "Production Association "Start" named after Michael V.Protsenko", Zarechny, Penza region, Russia

E-mail: yurii.spirichev@mail.ru

29.02.2020

It is shown that invariants and relativistically invariant laws of conservation of physical quantities in Minkowski space follow from 4-tensors of the second rank, which are four-dimensional derivatives of 4-vectors, tensor products of 4-vectors and inner products of 4-tensors of the second rank. Two forms of the system of equations of conservation laws for a number of physical quantities in Minkowski space are obtained. The four-dimensional law of conservation of energy-momentum combines the three-dimensional laws of conservation of energy, momentum and angular momentum. The equations of the four-dimensional laws of conservation of physical quantities in explicit or implicit form contain the wave part Based on a system of four-dimensional kinematic conservation equations, the reason for the stability of vortex rings in liquids and gases is explained.

**1 Introduction**
**2 Four-dimensional invariants of physical quantities**
**3 Four-dimensional laws of conservation of physical quantities**
**4 Conclusion**
**References**

**1 Introduction**

The laws of conservation of energy, momentum and angular momentum must be observed in any physical theory. For closed physical systems consisting of matter and fields, the law of conservation of energy and momentum has the form of 4-divergence of 4-tensor [1] $\partial^\mu T_{\mu\nu} = 0$, where $T_{\mu\nu}$ - full symmetric energy-momentum tensor (EMT) of a closed physical system. In the general case, the EMT may not be symmetrical, then $\partial^\mu T_{\mu\nu} + \partial^\nu T_{\mu\nu} = 0$. Asymmetric EMT can be decomposed into symmetric and antisymmetric 4-tensors:

$$\partial^\mu T_{(\mu\nu)} + \partial^\mu T_{[\mu\nu]} + \partial^\nu T_{(\mu\nu)} + \partial^\nu T_{[\mu\nu\}} = 0$$

Given that $\partial^\mu T_{(\mu\nu)} = \partial^\nu T_{(\mu\nu)}$, and $\partial^\mu T_{[\mu\nu]} = -\partial^\nu T_{[\mu\nu]}$, the equation of the laws of conservation of energy and momentum has the form of divergence of the symmetric part of the asymmetric TMT $\partial^\mu T_{(\mu\nu)} = 0$ [2].

Equations of the laws of conservation of energy and momentum can also be obtained on the basis of the variation principle, using the Lagrange formalism and invariants of the physical system. The invariants of physical systems are inextricably linked with their symmetries. According to G. Weil, symmetry should be understood as the invariability (invariance) of the properties of a physical object during transformations of a certain kind [3]. Thus, symmetry is a collection of invariant properties of an object. The connection between symmetries and conservation laws is established by the theorem of E. Noether [4], whose proof is based on the Lagrange formalism. Symmetries of space and time are the properties of their homogeneity and isotropy. The homogeneity of space and time means that there is no point with respect to which some distinguished symmetry exists and all points of space and time are equitable. According to E. Noether's theorem, the properties of the homogeneity of space and time correspond to independent laws of conservation of momentum and energy. Isotropy of space, i.e. the uniformity of its properties in all directions corresponds to the law of conservation of angular momentum. Exact conservation laws corresponding to the time isotropy property have not yet been found, although laws of conservation of parity of various physical quantities are associated with it. If space has the property of homogeneity, then it automatically has the property of isotropy, but not vice versa. It follows from this that the homogeneity of space is a special case of its isotropy. Therefore, the law of conservation of momentum for a homogeneous space should automatically include the law of conservation of angular momentum. This will be shown in section 3.

A change in views on space and time, as a single continuum that occurred at the beginning of the twentieth century in the works of A. Poincaré and G. Minkowski, led to a change in views on conservation laws. From the homogeneity of space-time follows a single law of conservation of energy-momentum. Space-time is considered Rimanov, but in small neighborhoods of a local point, in the absence of a strong gravitational field, it can be considered linear Minkowski space. However, in Minkowski space, depending on the accepted representation, either the time axis or spatial axes are imaginary [5], therefore it cannot be considered isotropic, therefore, it cannot be considered homogeneous either. This fact is expressed in physical observation of the "flow" of time and the local static nature of space. Nevertheless, a certain representation of four-dimensional physical quantities allows us to formally reduce Minkowski space-time to a homogeneous and isotropic form. Based on this representation of four-dimensional physical quantities, one can obtain their 4-invariants and four-dimensional conservation laws by a single method. The natural 4-invariants of physical quantities are traces of second rank 4-tensors constructed on the basis of these quantities. Such 4-tensors can be formed in three ways: four-dimensional differentiation of the 4-vector, tensor product of two 4-vectors and the inner product of two 4-tensors of the second rank. From these 4-tensors, in the form of their 4-divergences, one can obtain relativistically invariant conservation laws for the corresponding physical quantities.

In this paper, to obtain 4-invariants and four-dimensional conservation laws, vector quantities have a representation with imaginary spatial components and a real time component [5]. With this representation of 4-vectors, the geometry of Minkowski space-time corresponds to the four-dimensional Euclidean geometry in homogeneous and isotropic space. The four-dimensional density of a mechanical impulse will be described as a 4-vector $\mathbf{P}_\nu(c \cdot m, i \cdot \mathbf{p})$, where $m$ and $\mathbf{p}$ - mass density and particle momentum density vector, and c is the speed of light. The four-dimensional particle velocity will be described as a 4-vector $\mathbf{V}_\nu(c, i \cdot \mathbf{V})$, where $\mathbf{V}$ - particle velocity vector. The electromagnetic field (EMF) will be described using the electromagnetic 4-potential $\mathbf{A}_\nu(\varphi/c, i \cdot \mathbf{A})$, and charges and currents, in the form of a 4-vector current density $\mathbf{J}_\nu(c \cdot \rho, i \cdot \mathbf{J})$, where φ and $\mathbf{A}$ - are the scalar and vector potentials of the EMF, and ρ and $\mathbf{J}$ - are the charge density and the current density vector. For the accepted representation of the space-time geometry, the four-dimensional partial derivative operator has the form $\partial_\mu(\partial_t/c, i \cdot \nabla)$. For the accepted description of this operator and four-dimensional vector quantities, it is possible not to distinguish between covariant and contravariant indices.

The purpose of this article is to obtain 4-invariants and relativistically invariant four-dimensional conservation laws for a number of physical quantities in Minkowski space

**2 Four-dimensional invariants of physical quantities**

Usually, invariants of a physical system for use in the Lagrange formalism are obtained empirically based on general considerations [6]. In [7], a closed formalism is shown, including the Lagrange formalism and the method of mathematically obtaining the invariants of the energy of physical systems, in the form of traces of EMT. Traces of asymmetric and symmetric 4-tensors of the second rank are natural 4-invariants of physical quantities that are components of these 4-tensors. Since any asymmetric 4-tensor can be decomposed into symmetric and antisymmetric 4-tensors, its linear invariant follows from the symmetric part of the 4-tensor. Such tensors can be formed in three ways: four-dimensional differentiation of the 4-vector, tensor product of two 4-vectors and the inner product of two 4-tensors of the second rank. Next, we obtain 4-tensors of some four-dimensional physical quantities by the listed methods.

Will find the 4-derivative of the 4-velocity vector $\mathbf{V}_\nu(c, i\cdot\mathbf{V})$:

$$V_{\mu\nu} = \partial_\mu \mathbf{V}_\nu = \begin{pmatrix} 0 & \frac{1}{c}i\cdot\partial_t V_x & \frac{1}{c}i\cdot\partial_t V_y & \frac{1}{c}i\cdot\partial_t V_z \\ 0 & \partial_x V_x & \partial_x V_y & \partial_x V_z \\ 0 & \partial_y V_x & \partial_y V_y & \partial_y V_z \\ 0 & \partial_z V_x & \partial_z V_y & \partial_z V_z \end{pmatrix} \quad (1)$$

From this 4-tensor follows a 4-invariant of velocity $I_1 = \delta_{\mu\nu} V_{\mu\nu} = \nabla\cdot\mathbf{V} = \frac{1}{V}\frac{d\mathbf{V}}{dt}$.

Will find the 4-derivative of the 4-vector of momentum density $\mathbf{P}_\nu(c\cdot m, i\cdot\mathbf{p})$:

$$P_{\mu\nu} = \partial_\mu \mathbf{P}_\nu = \begin{pmatrix} \partial_t m & \frac{1}{c}i\cdot\partial_t P_x & \frac{1}{c}i\cdot\partial_t P_y & \frac{1}{c}i\cdot\partial_t P_z \\ -ci\cdot\partial_x m & \partial_x P_x & \partial_x P_y & \partial_x P_z \\ -ci\cdot\partial_y m & \partial_y P_x & \partial_y P_y & \partial_y P_z \\ -ci\cdot\partial_z m & \partial_z P_x & \partial_z P_y & \partial_z P_z \end{pmatrix} \quad (2)$$

From this 4-tensor follows the 4-invariant of the momentum density $I_2 = \delta_{\mu\nu} P_{\mu\nu} = \partial_t m + \nabla\cdot\mathbf{p}$, where $\delta_{\mu\nu}$ – Kronecker symbol.

Will find the 4-derivative of the 4-vector current density $\mathbf{J}_\nu(c\cdot\rho, i\cdot\mathbf{J})$:

$$J_{\mu\nu} = \partial_\mu \mathbf{J}_\nu = \begin{pmatrix} \partial_t \rho & \frac{1}{c}i\cdot\partial_t J_x & \frac{1}{c}i\cdot\partial_t J_y & \frac{1}{c}i\cdot\partial_t J_z \\ -ci\cdot\partial_x \rho & \partial_x J_x & \partial_x J_y & \partial_x J_z \\ -ci\cdot\partial_y \rho & \partial_y J_x & \partial_y J_y & \partial_y J_z \\ -ci\cdot\partial_z \rho & \partial_z J_x & \partial_z J_y & \partial_z J_z \end{pmatrix} \quad (3)$$

From this 4-tensor follows a 4-invariant current density $I_3 = \delta_{\mu\nu} J_{\mu\nu} = \partial_t \rho + \nabla\cdot\mathbf{J}$.

Will find the 4-derivative of the 4-potential $\mathbf{A}_\nu(\varphi/c, i\cdot\mathbf{A})$:

$$A_{\mu\nu} = \partial_\mu \mathbf{A}_\nu = \begin{pmatrix} \frac{1}{c^2}\partial_t\varphi & \frac{1}{c}i\cdot\partial_t A_x & \frac{1}{c}i\cdot\partial_t A_y & \frac{1}{c}i\cdot\partial_t A_z \\ -\frac{1}{c}i\cdot\partial_x\varphi & \partial_x A_x & \partial_x A_y & \partial_x A_z \\ -\frac{1}{c}i\cdot\partial_y\varphi & \partial_y A_x & \partial_y A_y & \partial_y A_z \\ -\frac{1}{c}i\cdot\partial_z\varphi & \partial_z A_x & \partial_z A_y & \partial_z A_z \end{pmatrix} \quad (4)$$

From this tensor follows the 4-invariant of the electromagnetic potential $I_4 = \delta_{\mu\nu} A_{\mu\nu} = \partial_t\varphi/c^2 + \nabla\cdot\mathbf{A}$.

Will find the tensor product of 4-velocity by 4-momentum:

$$T_{\mu\nu} = \mathbf{V}_\mu \otimes \mathbf{P}_\nu = \begin{pmatrix} m\cdot c^2 & i\cdot c\cdot p_x & i\cdot c\cdot p_y & i\cdot c\cdot p_z \\ i\cdot c\cdot p_x & -V_x\cdot p_x & -V_x\cdot p_y & -V_x\cdot p_z \\ i\cdot c\cdot p_y & -V_y\cdot p_x & -V_y\cdot p_y & -V_y\cdot p_z \\ i\cdot c\cdot p_z & -V_z\cdot p_x & -V_z\cdot p_y & -V_z\cdot p_z \end{pmatrix} \quad (5)$$

From this 4-tensor follows the 4-invariant of mechanical energy $I_5 = \delta_{\mu\nu} T_{\mu\nu} = m\cdot c^2 - \mathbf{p}\cdot\mathbf{V}$.

We find the tensor product of the 4-potential by the 4-vector of current density:

$$T_{\mu\nu} = \mathbf{A}_\mu \otimes \mathbf{J}_\nu = \begin{pmatrix} \rho \cdot \varphi & \frac{1}{c} i \cdot \varphi \cdot J_x & \frac{1}{c} i \cdot \varphi \cdot J_y & \frac{1}{c} i \cdot \varphi \cdot J_z \\ i \cdot c \cdot \rho \cdot A_x & -A_x \cdot J_x & -A_x \cdot J_y & -A_x \cdot J_z \\ i \cdot c \cdot \rho \cdot A_y & -A_y \cdot J_x & -A_y \cdot J_y & -A_y \cdot J_z \\ i \cdot c \cdot \rho \cdot A_z & -A_z \cdot J_x & -A_z \cdot J_y & -A_z \cdot J_z \end{pmatrix} \quad (6)$$

From this 4-tensor follows the 4-invariant of the interaction energy of electric charges with an EMF $I_6 = \delta_{\mu\nu} T_{\mu\nu} = \varphi \cdot \rho - \mathbf{A} \cdot \mathbf{J}$.

We give, obtained in [8], the internal product of the antisymmetric 4-tensor of the EMF and the antisymmetric 4-tensor of electromagnetic induction $T_{\nu\mu} = F_{[\nu\lambda]} \cdot f_{[\lambda\mu]}$. Since the expressions of the components of this 4-tensor are rather complicated, we write them in columns (in [8], the components of the 4-tensor are written in terms of the electric field **E**, magnetic field **H**, electric induction **D** and magnetic induction **B**):

$$\begin{aligned}
T_{00} &= E_x D_x + E_y D_y + E_z D_y & T_{01} &= i \cdot (E_y H_z - E_z H_y)/c \\
T_{11} &= E_x D_x - B_z H_z - B_y H_y & T_{02} &= i \cdot (E_z H_x - E_x H_z)/c \\
T_{22} &= E_y D_y - B_z H_z - B_x H_x & T_{03} &= i \cdot (E_x H_y - E_y H_x)/c \\
T_{33} &= E_z D_z - B_y H_y - B_x H_x & T_{10} &= ic(B_z D_y - B_y D_z) \\
T_{20} &= ic(B_x D_z - B_z D_x) & T_{30} &= ic(B_y D_x - B_x D_y) \\
T_{12} &= E_x D_y + B_y H_x & T_{13} &= E_x D_z + B_z H_x \\
T_{21} &= E_y D_x + B_x H_y & T_{23} &= E_y D_z + B_z H_y \\
T_{31} &= E_z D_x + B_x H_z & T_{32} &= E_z D_y + B_y H_z
\end{aligned} \quad (7)$$

This 4-tensor yields the well-known [6] 4-invariants of the EMF energy in a dielectric medium $I_7 = \delta_{\mu\nu}(F_{[\nu\lambda]} \cdot f_{[\lambda\mu]}) = \mathbf{E} \cdot \mathbf{D} - \mathbf{B} \cdot \mathbf{H}$ and in a vacuum $I_8 = \mathbf{E}^2/c^2 - \mathbf{B}^2$.

We briefly consider the obtained 4-tensors and 4-invariants. 4-tensors (1)-(4) describe, respectively, the rate of change in Minkowski space of four-dimensional velocity, momentum density, current density, electromagnetic potential. The 4-invariants of tensors (2)-(4) have the form of the canonical equations continuityof Euler, respectively, for the momentum density, current density, and vector potential. Invariant $I_4$ also has the form of the Lorentz gauge condition. 4-tensors (5)-(7) are EMT, respectively, for the dynamics of a continuous medium, for the interaction of an EMF with electric charges, for the antisymmetric part of a free EMF. Invariant $I_5$ is the Lagrangian of continuum mechanics, and $I_6$ is known as the generalized energy density of electromagnetic interaction or the Schwarzschild invariant [9] and is known in electrodynamics as the Lagrangian [6]. Invariants $I_7$ and $I_8$ are known energy invariants [6] of the antisymmetric part of the electromagnetic field. Tensors (1)-(7) are asymmetric and can be decomposed into symmetric and antisymmetric tensors. The symmetric parts of these 4-tensors have linear invariants that coincide with the invariants of the corresponding asymmetric tensors. Antisymmetric tensors do not have linear invariants, since their diagonal components are equal to zero. Such tensors, for example, the antisymmetric EMF 4-tensor, have quadratic invariants coinciding with the linear invariants I7 and $I_8$ of the EMT (7), which is an internal product of the EMF 4-tensors. Thus, to find the 4-invariant of a physical quantity, one of the listed methods is sufficient to find the corresponding 4-tensor and take its trace.

### 3 Four-dimensional laws of conservation of physical quantities

Three-dimensional laws of conservation of physical quantities are described by an equation relating the rate of change of their density over time in a certain volume with the divergence of the flux density of this physical quantity through the surface boundary that limits the volume. For the first time, an equation of this type was obtained by Euler for hydrodynamics and called the equation of continuity. Subsequently, similar equations were obtained for other physical quantities, for example, from the Maxwell equations, the continuity equation for the current density was obtained, which describes, as is believed, the law of conservation of electric charge.

The four-dimensional laws of conservation of physical quantities can be obtained from the corresponding symmetric 4-tensor of the second rank, in the form of its 4-divergence. In [2], from the tensor product of 4-vectors, 4-tensor (6) and four-dimensional conservation laws for the energy of interaction of charged particles with EMF are obtained, and in [8], from the 4-tensor (7) of the internal product of antisymmetric EMF tensors, four-dimensional laws of conservation of energy and momentum EMF. Запишем четырехмерный закон сохранения некоторой векторной физической величины в виде 4-дивергенции связанного с этой величиной симметричного 4-тензора We write the four-dimensional law of conservation of a certain vector physical quantity $\mathbf{N}_\nu(c \cdot k, i \cdot \mathbf{n})$ in the form of a 4-divergence of the symmetric 4-tensor associated with this quantity $N_{(\mu\nu)}$ for the accepted description of four-dimensional vector quantities, it is possible not to distinguish between covariant and contravariant indices):

$$\partial_\mu N_{(\mu\nu)} = 0 \qquad (8)$$

Eq. (8) can be written in expanded form as a system of equations:

$$\frac{1}{c^2}\partial_{tt}k - \Delta k + \frac{1}{c^2}\partial_t(\partial_t k + \nabla \cdot \mathbf{n}) = 0 \qquad (9)$$

$$\frac{1}{c^2}\partial_{tt}\mathbf{n} - \Delta \mathbf{n} - \nabla(\partial_t k + \nabla \cdot \mathbf{n}) = 0 \qquad (10)$$

These equations have a wave part in the form of the d'Alembert canonical wave equations for k and **n**. After transferring her to the right side of the equations, we obtain:

$$\frac{1}{c^2}\partial_t(\partial_t k + \nabla \cdot \mathbf{n}) = \Delta k - \frac{1}{c^2}\partial_{tt} k \qquad (11)$$

$$-\nabla(\partial_t k + \nabla \cdot \mathbf{n}) = \Delta \mathbf{n} - \frac{1}{c^2}\partial_{tt}\mathbf{n} \qquad (12)$$

Now the left-hand sides of these equations can be considered as a description of the wave sources described by the right-hand side. The expressions in brackets on the left side are in the form of the Euler continuity equation. Then Eqs. (9) and (10) can be considered as a system of complete four-dimensional equations of continuity of the 4-vector $\mathbf{N}_\nu(c \cdot k, i \cdot \mathbf{n})$. From Eqs. (11) and (12) follows the physical connection of three-dimensional continuity equations with wave equations. Since waves can exist independently of the source of their excitation, the left side of the equations can be equated to zero for non-zero components of the 4-vector k and $\mathbf{n}$. We write for this particular case Eqs. (11) and (12) in the form:

$$\partial_t(\partial_t k + \nabla \cdot \mathbf{n}) = 0 \qquad (13)$$

$$\nabla(\partial_t k + \nabla \cdot \mathbf{n}) = 0 \qquad (14)$$

After integrating these equations over time and space, we can write:

$$\partial_t k + \nabla \cdot \mathbf{n} = const(t, r) \qquad (15)$$

In the particular case when the constant in time and space on the right side of the equation is zero, this equation reduces to the canonical Euler continuity equation for the vector $\mathbf{n}$

$$\partial_t k + \nabla \cdot \mathbf{n} = 0 \qquad (16)$$

After substituting into the system of Eqs. (9) - (10) instead of the abstract 4-vector $\mathbf{N}_\nu(c \cdot k, i \cdot \mathbf{n})$, physical 4-vectors $\mathbf{P}_\nu(c \cdot m, i \cdot \mathbf{p})$, $\mathbf{J}_\nu(c \cdot \rho, i \cdot \mathbf{J})$, $\mathbf{A}_\nu(\varphi/c, i \cdot \mathbf{A})$, $\mathbf{V}_\nu(c, i \cdot \mathbf{V})$, we obtain, respectively, the continuity Eq. (16) in the form of well-known equations of the law of conservation of mass and charge and the Lorentz gauge condition, which can conditionally be considered the scalar potential conservation equation. Thus, the widely used continuity Eq. (16), which describes the conservation laws of many physical quantities, is a special case of the more general four-dimensional conservation law of physical quantities described by the system of Eqs. (9) - (10). Eq. (10) can be written in the form:

$$\frac{1}{c^2}\partial_{tt}\mathbf{n} - \nabla \partial_t k - \nabla(\nabla \cdot \mathbf{n}) - \Delta \mathbf{n} = 0 \text{ ore } \frac{1}{c^2}\partial_{tt}\mathbf{n} - \nabla \partial_t k - 2\nabla(\nabla \cdot \mathbf{n}) + \nabla \times \nabla \times \mathbf{n} = 0 \qquad (17)$$

This form corresponds to the equation known in the theory of continuous media as the Lame equation of motion [10] or the dynamic Euler equation [11]. This equation can be considered a dynamic continuity equation for the three-dimensional vector $\mathbf{n}$.

We take the time derivative of Eq. (12) and, replacing the brackets in it taken from Eq. (11), we obtain the vector wave equation:

$$\frac{1}{c^2}\partial_t(\Delta \mathbf{n} - \frac{1}{c^2}\partial_{tt}\mathbf{n}) + \nabla(\Delta k - \frac{1}{c^2}\partial_{tt}k) = 0 \qquad \text{ore} \qquad (\Delta - \frac{1}{c^2}\partial_{tt})(\frac{1}{c^2}\partial_t\mathbf{n} + \nabla k) = 0 \quad (18)$$

This equation combines Eqs. (9) and (10) and describes the wave law of conservation of a 4-vector in Minkowski space. From Eqs. (11), (12) and (18) also follows the equation:

$$\partial_t \nabla(\partial_t k + \nabla \cdot \mathbf{n}) = 0 \tag{19}$$

This equation can be written as:

$$\partial_t(\nabla \partial_t k + \nabla \times \nabla \times \mathbf{n} + \Delta \mathbf{n}) = 0 \tag{20}$$

After integration over time, we obtain:

$$\partial_t \nabla k + \nabla \times \nabla \times \mathbf{n} + \Delta \mathbf{n} = const(t) \tag{21}$$

The right side of this equation is a time constant vector. Thus, the system of equations of conservation laws (9) - (10) can be written in the form of a system consisting of wave Eqs. (18) and (19). This system emphasizes the inextricable relationship in the four-dimensional laws of conservation of wave equations with continuity equations.

We consider the system of Eqs. (9) - (10) of the conservation laws for the momentum density of a continuous medium in the absence of interaction of a mechanical system with other physical systems. Substituting a 4-vector $\mathbf{P}_\nu(c \cdot m, i \cdot \mathbf{p})$ Eqs. (9) and (10), we obtain conservation equations for four-dimensional mechanics of a continuous medium:

$$\frac{1}{c^2}\partial_{tt}m - \Delta m + \frac{1}{c^2}\partial_t(\partial_t m + \nabla \cdot \mathbf{p}) = 0 \tag{22}$$

$$\frac{1}{c^2}\partial_{tt}\mathbf{p} - \Delta \mathbf{p} - \nabla(\partial_t m + \nabla \cdot \mathbf{p}) = 0 \tag{23}$$

Substituting the 4-vector $\mathbf{P}_\nu(c \cdot m, i \cdot \mathbf{p})$ into Eqs. (18) - (21), we obtain the equations:

$$(\Delta - \frac{1}{c^2}\partial_{tt})(\frac{1}{c^2}\partial_t \mathbf{p} + \nabla m) = 0 \tag{24}$$

$$\partial_t \nabla(\partial_t m + \nabla \cdot \mathbf{p}) = \partial_t(\partial_t \nabla m + \nabla \times \nabla \times \mathbf{p} + \Delta \mathbf{p}) = 0 \quad \text{ore} \quad \partial_t \nabla m + \nabla \times \nabla \times \mathbf{p} + \Delta \mathbf{p} = const(t) \tag{25}$$

The last conservation equation contains a term $\nabla \times \nabla \times \mathbf{p}$ describing the rotational motion of the medium, i.e. contains the angular momentum, which corresponds to the homogeneity and isotropy of space. Conservation Eq. (25) confirms the statement made earlier that since the homogeneity of space is a special case of its isotropy, the complete equation of the law of conservation of momentum must include the law of conservation of angular momentum.

Consider the system of Eqs. (9) - (10) of conservation laws for the 4-vector velocity in the form of equations of four-dimensional kinematics of a continuous medium:

$$\partial_t(\nabla \cdot \mathbf{V}) = \partial_t(\frac{1}{\mathbf{V}}\frac{d\mathbf{V}}{dt}) = 0 \text{ или } \nabla \cdot \mathbf{V} = \frac{1}{\mathbf{V}}\frac{d\mathbf{V}}{dt} = const(t) \tag{26}$$

$$\nabla(\nabla \cdot \mathbf{V}) = \frac{1}{c^2}\partial_{tt}\mathbf{V} - \Delta\mathbf{V} \quad \text{ore} \quad 2\nabla \times \mathbf{\Omega} + \Delta\mathbf{V} = \frac{1}{c^2}\partial_{tt}\mathbf{V} - \Delta\mathbf{V} \tag{27}$$

The left side of this equation includes the rotor of the angular velocity vector $\Omega$. This conservation equation can be interpreted as a description of the kinematics of toroidal vortex structures, for example, widely known vortex rings in liquids and gases. Vortex rings are stable dynamic structures moving in gas and liquid. Their stability is explained by the fact that they are described by the kinematic conservation equation. The kinematic Eqs. (26) - (27) are associated with the dynamics Eqs. (22) - (25) for the momentum density.

Substituting the velocity 4-vector $\mathbf{V}_\nu(c, i \cdot \mathbf{V})$ into the system of Eqs. (18) and (19), we obtain the system of equations of four-dimensional kinematics of a continuous medium in the form:

$$\Delta(\partial_t \mathbf{V}) - \frac{1}{c^2}\partial_{tt}(\partial_t \mathbf{V}) = 0 \tag{28}$$

$$\nabla(\nabla \cdot \partial_t \mathbf{V}) = 0 \quad \text{ore} \quad 2\nabla \times \partial_t \mathbf{\Omega} + \Delta \partial_t \mathbf{V} = 0 \quad \text{ore} \quad \nabla \times \mathbf{\Omega} + \Delta \mathbf{V} = const(t) \tag{29}$$

The conservation in time of the left side of this equation, includes the rotor of angular velocity, explains the stability of toroidal vortex rings in liquids and gases.

We consider the conservation law (10) for the tensor product of 4-vectors of velocity $\mathbf{V}_\nu(c, i \cdot \mathbf{V})$ and momentum density $\mathbf{P}_\nu(c \cdot m, i \cdot \mathbf{p})$ (5). This tensor is a tensor of mechanical energy-momentum and is symmetric. It follows the system of equations for the conservation of energy and momentum density for a neutral isotropic continuous medium, in the absence of external forces:

$$\partial_t m + \nabla \cdot \mathbf{p} = 0 \tag{30}$$

$$\partial_t \mathbf{p} + \nabla(\mathbf{V} \cdot \mathbf{p}) + \mathbf{p}(\nabla \cdot \mathbf{V}) + \mathbf{V}(\nabla \cdot \mathbf{p}) - \mathbf{V} \times \nabla \times \mathbf{p} - \mathbf{p} \times \nabla \times \mathbf{V} = 0 \tag{31}$$

Substituting Eq. (30) in Eq. (31) we get one equation:

$$m\partial_t \mathbf{V} + \nabla(\mathbf{V} \cdot \mathbf{p}) + \mathbf{p}(\nabla \cdot \mathbf{V}) - \mathbf{V} \times \nabla \times \mathbf{p} - \mathbf{p} \times \nabla \times \mathbf{V} = 0 \tag{32}$$

The first term of Eq. (31) includes the density of Newton's inertial force. The last two terms represent the density of the centrifugal inertia force of rotation of the medium and the moment of momentum enters into them. Therefore, the system of conservation equations (30) - (31) simultaneously describes the conservation of momentum and angular momentum. It follows that in Minkowski space the individual three-dimensional laws of conservation of energy, momentum and angular momentum are combined into one law of conservation of energy-momentum.

We consider the system of Eqs. (9) - (10) of conservation laws for a charge-free EMF. When substituting the electromagnetic 4-potential $\mathbf{A}_\nu(\varphi/c, i \cdot \mathbf{A})$ in them, we obtain the equations of the laws of conservation of EMF in a vacuum:

$$\frac{1}{c^2}\partial_{tt}\varphi - \Delta\varphi + \partial_t(\frac{1}{c^2}\partial_t\varphi + \nabla\cdot\mathbf{A}) = 0 \qquad (33)$$

$$\frac{1}{c^2}\partial_{tt}\mathbf{A} - \Delta\mathbf{A} - \nabla(\frac{1}{c^2}\partial_t\varphi + \nabla\cdot\mathbf{A}) = 0 \qquad (34)$$

Substituting the electromagnetic 4-potential $\mathbf{A}_\nu(\varphi/c, i\cdot\mathbf{A})$ into the wave Eq. (20), we obtain:

$$\partial_t(\Delta\mathbf{A} - \frac{1}{c^2}\partial_{tt}\mathbf{A}) + \nabla(\Delta\varphi - \frac{1}{c^2}\partial_{tt}\varphi) = 0 \qquad (35)$$

Having selected the wave operator, we bring this equation to the form:

$$(\Delta - \frac{1}{c^2}\partial_{tt})(-\partial_t\mathbf{A} - \nabla\varphi) = \Delta\mathbf{E} - \frac{1}{c^2}\partial_{tt}\mathbf{E} = 0 \qquad (36)$$

It follows that one of the equations for the conservation of electromagnetic fields is a canonical wave equation for an electric field in vacuum, well known in electrodynamics. The wave Eq. (36) includes not only the vortex part of the electric field, but also its potential part. Substituting the electromagnetic 4-potential $\mathbf{A}_\nu(\varphi/c, i\cdot\mathbf{A})$ into the Eqs. (19) - (21), we obtain:

$$\partial_t\nabla(\partial_t\varphi/c^2 + \nabla\cdot\mathbf{A}) = \partial_t(\partial_t\nabla\varphi/c^2 + \nabla(\nabla\cdot\mathbf{A})) = \partial_t(\partial_t\nabla\varphi/c^2 + \nabla\times\nabla\times\mathbf{A} + \Delta\mathbf{A}) = 0 \qquad (37)$$

or, after integration over time, we obtain:

$$\partial_t\nabla\varphi/c^2 + \nabla\times\nabla\times\mathbf{A} + \Delta\mathbf{A} = \mathbf{K} \qquad (38)$$

where $\mathbf{K}$ - is a vector constant in time. We rewrite Eq. (38) in the form

$$\nabla\times\mathbf{B} + \Delta\mathbf{A} = \partial_t\mathbf{E}/c^2 + \mathbf{K} \qquad (39)$$

Here $\mathbf{E}$ - is the potential electric field. We take a vector $\mathbf{K} = \mu_0\mathbf{J}$, and we obtain Eq. (39) in the form:

$$\nabla\times\mathbf{B} + \Delta\mathbf{A} = \partial_t\mathbf{E}/c^2 + \mu_0\mathbf{J} \qquad (40)$$

This equation is similar to the Maxwell equation, but has differences: it contains an additional term $\Delta\mathbf{A}$, the potential electric field $\mathbf{E}$, and the current density $\mathbf{J}$ is a constant. Therefore, Eq. (40) is valid only for direct current. If the current density is a variable, then Eq. (40) should be supplemented by wave terms, including a vortex electric field, i.e. in this case, it is necessary to apply the complete system of Eqs. (33) - (34).

We consider the system of Eqs. (9) - (10) of the conservation laws for the current density. Substituting 4-vector $\mathbf{J}_\nu(c\cdot\rho, i\cdot\mathbf{J})$ into them, we obtain the conservation equations:

$$\frac{1}{c^2}\partial_{tt}\rho - \Delta\rho + \frac{1}{c^2}\partial_t(\partial_t\rho + \nabla\cdot\mathbf{J}) = 0 \qquad (41)$$

$$\frac{1}{c^2}\partial_{tt}\mathbf{J} - \Delta\mathbf{J} - \nabla(\partial_t\rho + \nabla\cdot\mathbf{J}) = 0 \qquad (42)$$

Substituting the 4-vector $\mathbf{J}_\nu(c\cdot\rho, i\cdot\mathbf{J})$ into Eqs. (19) - (21), we obtain the system of equations:

$$(\Delta - \frac{1}{c^2}\partial_{tt})(\frac{1}{c^2}\partial_t \mathbf{J} + \nabla\rho) = 0 \qquad (43)$$

$$\partial_t \nabla(\partial_t \rho + \nabla \cdot \mathbf{J}) = \partial_t(\partial_t \nabla\rho + \nabla\times\nabla\times\mathbf{J} + \Delta\mathbf{J}) = 0 \quad \text{ore} \quad \nabla\partial_t\rho + \nabla\times\nabla\times\mathbf{J} + \Delta\mathbf{J} = const(t) \qquad (44)$$

It should be taken into account that electric charges have mass, therefore Eqs. (41) - (44) should be supplemented by similar Eqs. (22) - (25) for the 4-vector $\mathbf{P}_\nu(c\cdot m, i\cdot\mathbf{p})$ of momentum density or an equation describing the relationship between the charge density and the mass density should be added. In addition, moving electric charges are sources of EMF and interact with each other therefore, these equations need to be supplemented by the equations of conservation of free EMF (33) and (34), as well as the equations of interaction of electric charges with EMF. The equations of the laws of conservation of energy of interaction of electric charges with EMF will be given below. Thus, the system of conservation Eqs. (41) - (42), without taking into account other conservation laws, is valid only for massless electric charges that do not interact with each other and do not emit EMF.

We obtain four-dimensional conservation laws resulting from the 4-tensor formed by the tensor product of the 4-potential by the 4-vector of current density, i.e. 4-tensor (6). This tensor is asymmetric. We find its symmetric part and substitute the resulting symmetric 4-tensor in Eq. (10).

$$\partial_\mu T_{(\mu\nu)} = \partial_\mu \begin{pmatrix} 2\rho\cdot\varphi & \frac{1}{c}i\cdot\varphi\cdot J_x + i\cdot c\cdot\rho\cdot A_x & \frac{1}{c}i\cdot\varphi\cdot J_y + i\cdot c\cdot\rho\cdot A_y & \frac{1}{c}i\cdot\varphi\cdot J_z + i\cdot c\cdot\rho\cdot A_z \\ \frac{1}{c}i\cdot\varphi\cdot J_x + i\cdot c\cdot\rho\cdot A_x & -2A_x\cdot J_x & -A_x\cdot J_y - A_y\cdot J_x & -A_x\cdot J_z - A_z\cdot J_x \\ \frac{1}{c}i\cdot\varphi\cdot J_y + i\cdot c\cdot\rho\cdot A_y & -A_x\cdot J_y - A_y\cdot J_x & -2A_y\cdot J_y & -A_y\cdot J_z - A_z\cdot J_y \\ \frac{1}{c}i\cdot\varphi\cdot J_z + i\cdot c\cdot\rho\cdot A_z & -A_x\cdot J_z - A_z\cdot J_x & -A_y\cdot J_z - A_z\cdot J_y & -2A_z\cdot J_z \end{pmatrix} = 0$$

It should be noted that Eq. (10) is written for a closed physical system, so the electric charges in this example will be considered massless particles. For massive particles, this 4-tensor must be supplemented with a similar 4-tensor of mechanical energy-and-pulse (5). This is done in [2]. In addition, it must be taken into account that electric charges have their own EMF. In this example, we assume that the electric charges are in a strong external EMF, significantly exceeding the intrinsic EMF of the charges. Thus, this EMT does not take into account the energy of interaction of electric charges through its own EMF, as well as the energy of a free EMF emitted by electric charges in the form of electromagnetic waves. To take these factors into account, this system of equations needs to be supplemented by the EMF conservation Eqs. (33) - (34) and equations describing charges and currents as EMF sources. We write the system of equations for the conservation of energy and momentum of interaction of electric charges with EMF in expanded form:

$$2\partial_t(\varphi\cdot\rho) + \nabla\cdot(c^2\cdot\rho\cdot\mathbf{A} + \varphi\cdot\mathbf{J}) = 0 \qquad (45)$$

$$\frac{1}{c^2}\partial_t(\varphi\cdot\mathbf{J}) + \partial_t(\rho\cdot\mathbf{A}) + \nabla(\mathbf{A}\cdot\mathbf{J}) + \mathbf{J}(\nabla\cdot\mathbf{A}) + \mathbf{A}(\nabla\cdot\mathbf{J}) - \mathbf{A}\times\nabla\times\mathbf{J} - \mathbf{J}\times\nabla\times\mathbf{A} = 0 \qquad (46)$$

In [8], a system of equations for the conservation of energy and momentum of the EMF arising from the tensor (7), which is the internal product of the antisymmetric 4-tensor of the EMF and the antisymmetric 4-tensor of electromagnetic induction, is obtained $T_{\nu\mu} = F_{[\nu\lambda]} \cdot f_{[\lambda\mu]}$:

$$\frac{1}{c^2}\partial_t \mathbf{E}^2 + \nabla \cdot (\mathbf{E} \times \mathbf{B}) = 0 \quad \text{and} \quad \partial_t (\mathbf{E} \times \mathbf{B}) + \nabla \mathbf{E}^2 = 0$$

Further from these equations, wave equations for the energy and momentum of the EMF in vacuum follow:

$$\frac{1}{c^2}\partial_{tt}\mathbf{E}^2 - \Delta \mathbf{E}^2 = 0 \quad \text{and} \quad \frac{1}{c^2}\partial_{tt}(\mathbf{E}\times\mathbf{B}) - \nabla(\nabla \cdot (\mathbf{E}\times\mathbf{B})) = \frac{1}{c^2}\partial_{tt}(\mathbf{E}\times\mathbf{B}) - \nabla \times \nabla \times (\mathbf{E}\times\mathbf{B}) - \Delta(\mathbf{E}\times\mathbf{B}) = 0$$

The wave equation for an electromagnetic pulse describes the transfer of momentum and angular momentum.

**4 Conclusion**

The consistent application of the theory of four-dimensional space-time Punkare-Minkowski allows using the unified method to obtain 4-invariants of many physical quantities and relativistically invariant systems of equations that describe the four-dimensional conservation laws of these quantities

The natural 4-invariants of physical quantities are traces of 4-tensors of the second rank, the components of which include these quantities. These 4-tensors are four-dimensional derivatives of 4-vectors of physical quantities, tensor products of two 4-vectors and internal products of two 4-tensors of the second rank. Thus, to obtain a 4-invariant of a physical quantity, it is necessary to construct a 4-vector whose components this physical quantity will be. Then find the four-dimensional derivative of this 4-vector, which is a 4-tensor of the second rank. The trace of this 4-tensor is a 4-invariant. In the other case, it is necessary to construct two 4-vectors physically interconnected, and find their tensor product. The trace of this 4-tensor is a 4-invariant. In the third case, it is necessary to construct two 4-tensors of the second rank with components that are a physical quantity. Then find the inner product of these 4-tensors. The trace of this 4-tensors is a 4-invariant, which includes a physical quantity.

Relativistically invariant four-dimensional laws of conservation of physical quantities follow from symmetric 4-tensors or symmetric parts of asymmetric 4-tensors in the form of their 4-divergences. The equations of the four-dimensional laws of conservation of physical quantities in explicit or implicit form contain the wave part.

The four-dimensional kinematic conservation equations describe the kinematics of toroidal vortex structures in a continuous medium. Such structures are known as stable vortex rings. The stability of their existence in liquids and gases is explained by the fact that they are described by kinematic and dynamic equations of conservation in a continuous medium of Minkowski space. It also follows from this that vortex rings are fundamental dynamic formations associated with the geometry of Minkowski space.

In Minkowski space, individual three-dimensional laws of conservation of energy, momentum and angular momentum are combined into one law of conservation of energy-momentum.